\documentclass[conference]{IEEEtran}
\IEEEoverridecommandlockouts

\usepackage{cite}
\usepackage{bm}
\usepackage{amsmath,amssymb,amsfonts}
\usepackage{amsthm}        % for \newtheorem
\usepackage{algorithmic}
\usepackage{graphicx}
\usepackage{textcomp}
\usepackage[dvipsnames]{xcolor}
\usepackage{enumitem}
\usepackage{dblfloatfix}
\usepackage{float}
\usepackage{adjustbox}
\usepackage{booktabs}
\usepackage{tabularx}
\usepackage{array}
\usepackage{multirow}
\usepackage{makecell}
\usepackage{tikz}
\usetikzlibrary{shapes,arrows}
\usepackage{mathtools}
\usepackage[normalem]{ulem}
\usepackage{subcaption}
\usepackage{caption}
\usepackage{siunitx}

\usepackage[T1]{fontenc}
\usepackage{newtxtext,newtxmath}

\usepackage{bbm} % for \mathbbm (indicator)
\usepackage[hidelinks,bookmarks=false]{hyperref} % keep hyperref near the end
\usepackage{tikz}
\usetikzlibrary{arrows.meta,calc,positioning}
\usetikzlibrary{calc}
\newcolumntype{K}{>{\raggedright\arraybackslash}m{2.2 cm}}
\newcolumntype{E}{>{\centering\arraybackslash}m{2.3 cm}}
\newcolumntype{Y}{>{\centering\arraybackslash}m{4.2 cm}}
\newcolumntype{C}{>{\centering\arraybackslash}m{3.1 cm}}
\newcolumntype{d}{>{\centering\arraybackslash}m{1.35 cm}}
\newcolumntype{a}{>{\centering\arraybackslash}m{1.2 cm}}
\newcolumntype{e}{>{\centering\arraybackslash}m{1.35 cm}}
\newcolumntype{D}{>{\centering\arraybackslash}m{3 cm}}
\newcolumntype{z}{>{\centering\arraybackslash}m{0.7 cm}}
\newcolumntype{L}{>{\centering\arraybackslash}m{1 cm}}
\newcolumntype{x}{>{\centering\arraybackslash}m{0.75 cm}}
\newcolumntype{P}{>{\centering\arraybackslash}m{1.5 cm}}

\def\BibTeX{{\rm B\kern-.05em{\sc i\kern-.025em b}\kern-.08em
    T\kern-.1667em\lower.7ex\hbox{E}\kern-.125emX}}

\newcommand{\bsym}{\boldsymbol}

\def\beq{\begin{equation}}
\def\eeq{\end{equation}}
\def\beqa{\begin{align}}
\def\eeqa{\end{align}}
\def\beqan{\begin{align*}}
\def\eeqan{\end{align*}}
\def\nn{\nonumber}

\newcommand{\T}{^{\mathsf T}}
\newcommand{\Her}{^{\mathsf H}}
\newcommand{\herm}{^{\text{\sf H}}}

\newcommand{\hypcmp}[2]{\underset{#2}{\overset{#1}{\gtrless}}}

\def\C{{\mathbb{C}}}

\def\argmin{\mathop{\mathrm{arg\,min}}}

\def\Tr{\mathop{\mathrm{Tr}}}

\def\Exp{\mathbb{E}}

\newcommand{\CN}{\mathcal{CN}}

\newcommand{\subsf}{\sf \scriptscriptstyle}
\newcommand{\mc}{\mathcal}

\def\INR  {{\mathsf{INR}}}

\def\tm1{t\! - \! 1}
\def\tp1{t\! + \! 1}

\def\bbf{\mathbf{b}}

\def\ebf{\mathbf{e}}

\def\hbf{\mathbf{h}}

\def\qbf{\mathbf{q}}

\def\sbf{\mathbf{s}}
\def\ubf{\mathbf{u}}
\def\vbf{\mathbf{v}}
\def\wbf{\mathbf{w}}

\def\Ibf{\mathbf{I}}

\def\Qbf{\mathbf{Q}}
\def\Rbf{\mathbf{R}}

\def\Ubf{\mathbf{U}}

\def\Wbf{\mathbf{W}}

\def\Ybf{\mathbf{Y}}

\newcommand{\RX}[1]{\mathrm{RX}_{#1}}

\begin{document}

\title{Noncoherent Detection and Interference Nulling \\ for Terrestrial-Satellite Downlink \\  Coexistence in the Upper Mid-Band}
%Upper Mid-Band Interference Nulling

    \vspace{-15pt}

    \author{
    \IEEEauthorblockN{Shizhen Jia\IEEEauthorrefmark{1}, C. Nicolas Barati\IEEEauthorrefmark{2},  Marco Mezzavilla\IEEEauthorrefmark{3} and Sundeep Rangan\IEEEauthorrefmark{1}}

    \IEEEauthorblockA{\IEEEauthorrefmark{1}NYU WIRELESS, New York University, Brooklyn, NY, USA 
    }
    \IEEEauthorblockA{\IEEEauthorrefmark{2}Old Dominion University, Norfolk, VA, USA
        }
    \IEEEauthorblockA{\IEEEauthorrefmark{3}Dipartimento di Elettronica, Informazione e Bioingegneria (DEIB), Politecnico di Milano, Milan, Italy  \\
    \{s.jia, srangan\}@nyu.edu
    }
    \thanks{The appendix referenced in the paper is available at https://github.com/Shizhen-Jia/NTN-NULLING-NONCOH/Appendix.pdf.}
    \vspace{-22pt}
}
\bstctlcite{IEEEexample:BSTcontrol}
\maketitle

% ===== Top banner above title (no layout impact) =====
\begin{tikzpicture}[remember picture,overlay]
\node[
    anchor=north,
    align=center,
    font=\footnotesize\itshape,
    text width=\paperwidth
] at ([yshift=-0.25in]current page.north)
{S.~Jia, C.~N.~Barati, M.~Mezzavilla, and S.~Rangan, ``Noncoherent Detection and Interference Nulling for Terrestrial-Satellite Downlink Coexistence in the Upper Mid-Band,'' to appear in \textit{IEEE 27th International Workshop on Signal Processing Advances in Wireless Communications (SPAWC)}, 2026.};
\end{tikzpicture}

% ===== arXiv accepted-paper banner: placed above title, no page-length impact =====
% \begin{tikzpicture}[remember picture,overlay]
% \node[
%     anchor=north,
%     align=center,
%     font=\footnotesize\itshape,
%     text width=\paperwidth
% ] at ([yshift=-0.12in]current page.north)
% {Accepted for presentation at the 2026 IEEE International Workshop on Signal Processing Advances in Wireless Communications (SPAWC 2026).};
% \end{tikzpicture}

\begin{abstract}
Terrestrial--satellite coexistence in the upper mid-band is challenging when a terrestrial base station has limited prior information about non-terrestrial receivers and their uplink transmissions. This paper studies noncoherent victim sensing and interference nulling, where uplink sensing snapshots are observed while the transmitted waveform is treated as unknown. For a single victim, we show that the generalized likelihood-ratio test reduces to a principal-eigenvector estimator of the sample covariance. For multiple victims, we combine MDL-based model-order selection with MUSIC to recover anonymous direction and power information that is sufficient for beam design without pilot knowledge or user identities. These estimates are then used in a nulling beamformer that preserves the intended terrestrial link while reducing leakage toward detected victims. We further analyze the single-victim estimator in the large-matrix regime and show that estimation accuracy improves with sensing SNR, which reveals the observed interplay between path loss and estimation quality. Site-specific ray-tracing results show significant reduction of NTN INR with only modest degradation of TN SINR.
% Terrestrial--satellite coexistence in the upper mid-band is difficult when a terrestrial base station does not know which non-terrestrial receivers are active, where they are located, or what waveform they transmit. This paper studies noncoherent victim sensing and nulling, where victim uplink snapshots are observed but the temporal waveform is treated as an unknown nuisance parameter. For a single victim, we show that the generalized likelihood-ratio test (GLRT) reduces to a principal-eigenvector estimator of the sample covariance. For multiple victims, we combine MDL-based model-order selection with MUSIC to recover anonymous direction/power tuples that are sufficient for beam design without pilot knowledge or user identities. These estimates are then used in a nulling beamformer that preserves the intended terrestrial link while suppressing leakage toward detected victims. We further analyze the single-victim GLRT in the large-matrix regime and show that the overlap between the estimated and true spatial signatures improves as $1-\eta=O(\gamma_{\mathrm S}^{-1})$, which explains the observed residual-leakage trend after nulling. Site-specific ray-tracing results show significant reduction of NTN INR with only modest degradation of TN SINR.
\end{abstract}


\section{Introduction}

Satellite and other non-terrestrial networks (NTN) are becoming increasingly intertwined with terrestrial cellular networks, making terrestrial--satellite coexistence a central problem for future wireless systems. The challenge is particularly acute in the upper mid-band (7--\SI{24}{GHz}), which offers much more bandwidth than sub-\SI{7}{GHz} cellular spectrum while retaining more favorable propagation than mmWave bands~\cite{kang2024cellular}. Since this range overlaps with spectrum used by satellite and other incumbent non-terrestrial systems, terrestrial operation in this band requires effective mechanisms to limit harmful downlink interference~\cite{fcc,ntia}.

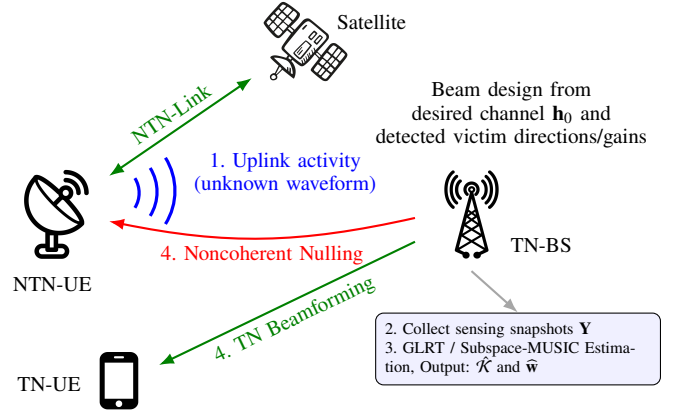
\begin{figure}[t]
    \centering
    \begin{tikzpicture}[
        scale=0.88,
        transform shape,
        >={Latex[length=2mm]},
        every node/.style={font=\small}
        ]

        %--------------------------------------------------
        % Nodes
        %--------------------------------------------------
        \node (sat) at (4.8,5.6)
            {\includegraphics[width=1.35cm]{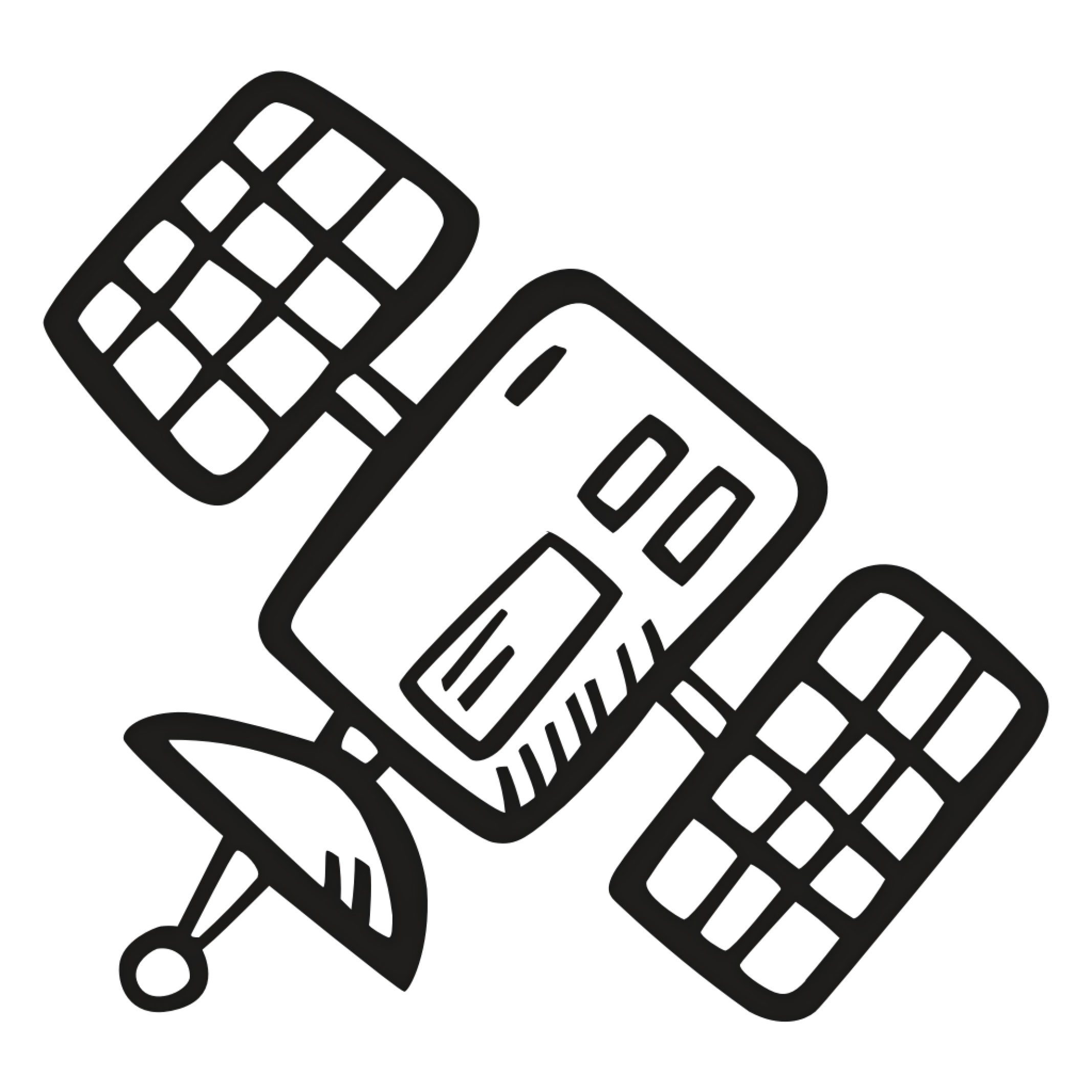}};
        \node at (5.8,5.9) {Satellite};

        \node (ntn) at (1.0,3.0)
            {\includegraphics[width=1.45cm]{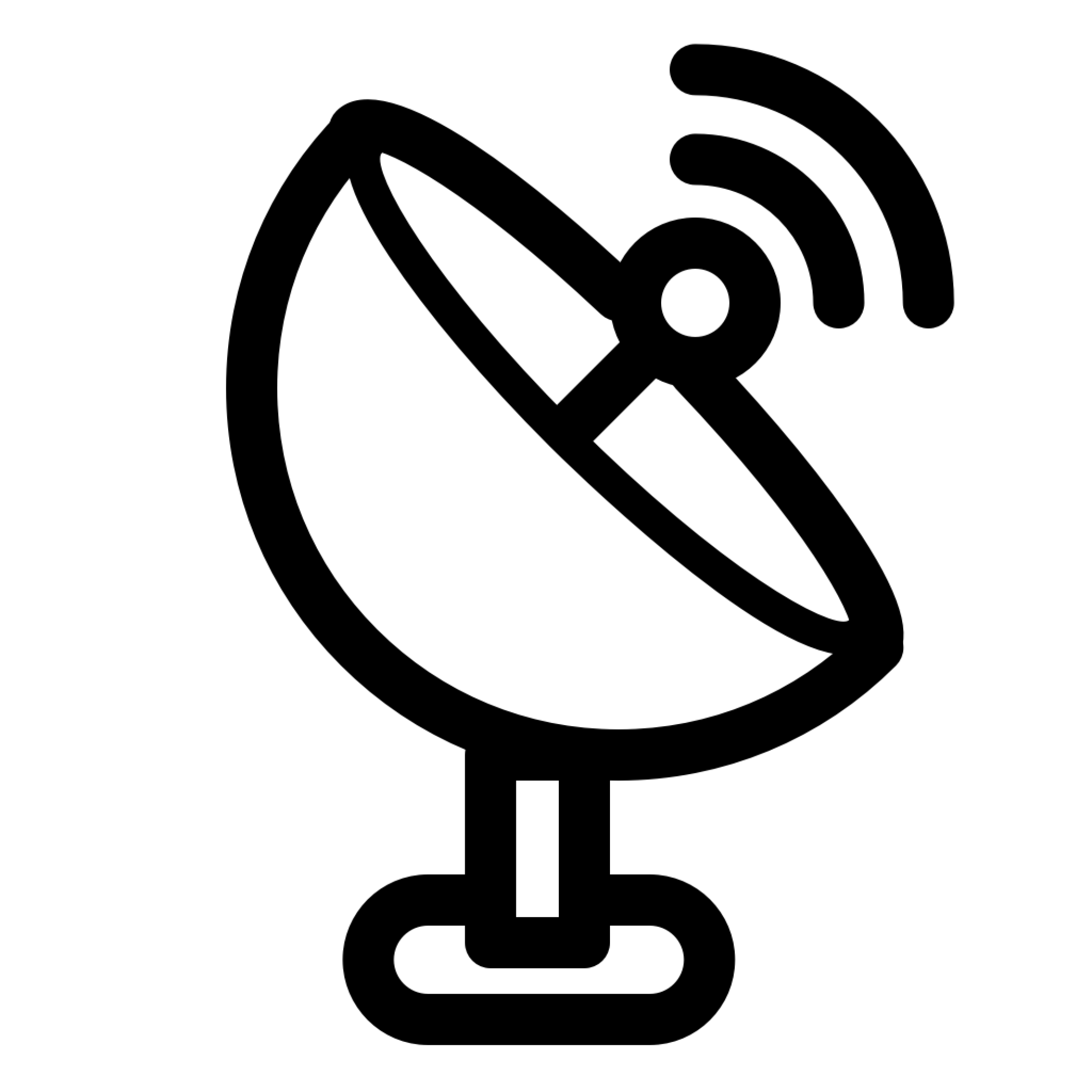}};
        \node at (1.0,1.95) {NTN-UE};

        \node (bs) at (7.3,3.0)
            {\includegraphics[width=1.45cm]{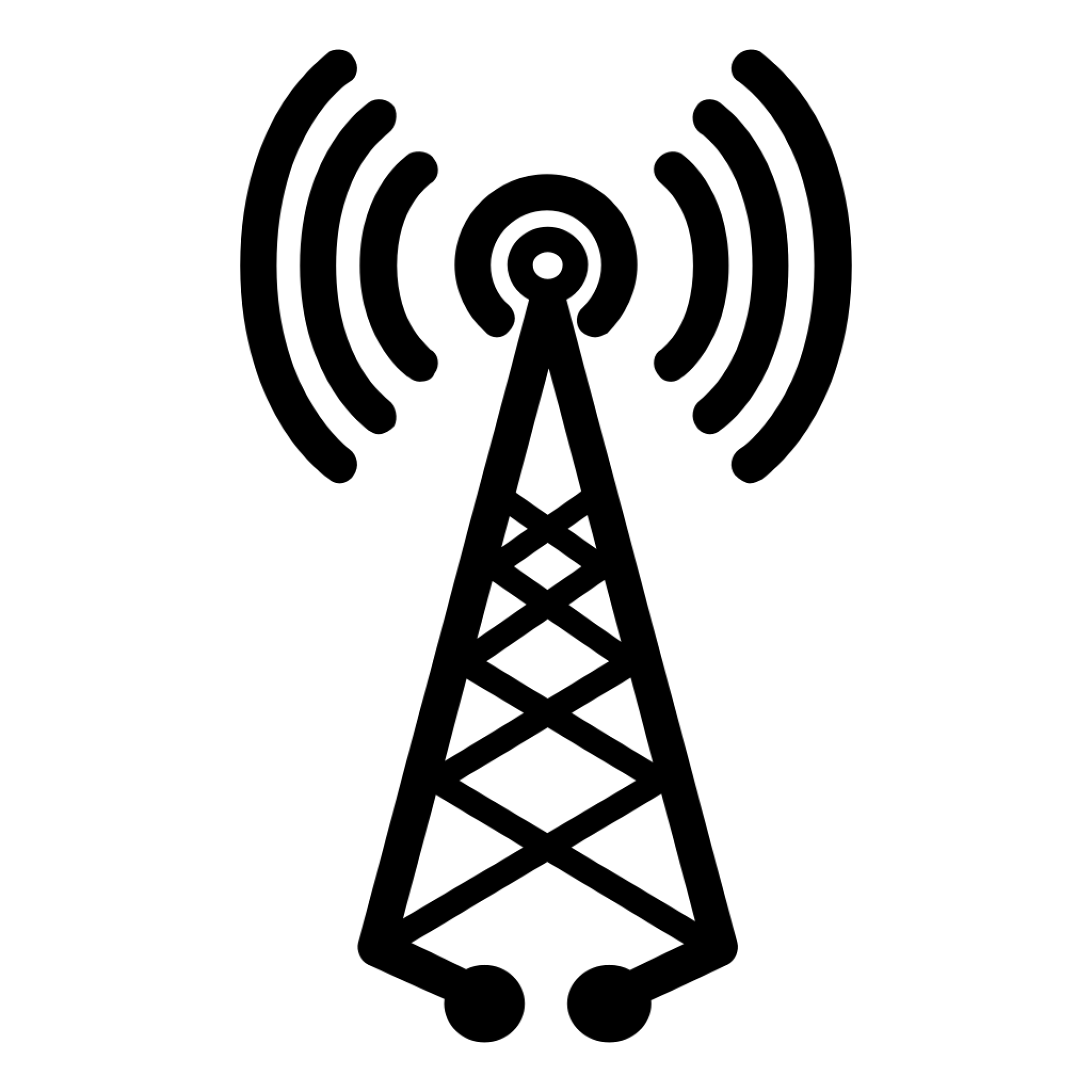}};
        \node at (8.35,2.55) {TN-BS};

        \node (tn) at (2.0,0.45)
            {\includegraphics[width=0.95cm]{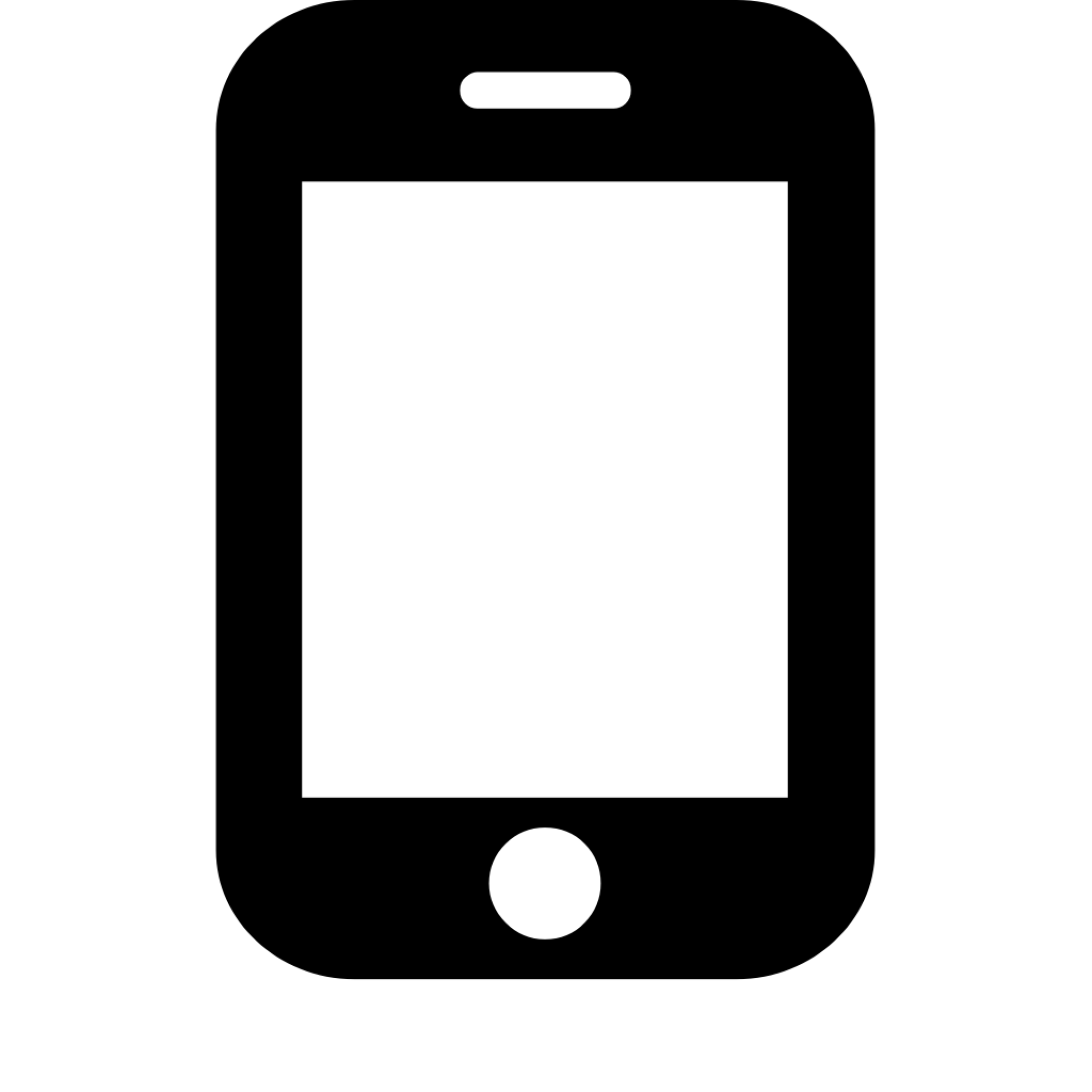}};
        \node at (0.95,0.45) {TN-UE};

        %--------------------------------------------------
        % Processing box
        %--------------------------------------------------
        \node[
            draw,
            rounded corners=3pt,
            fill=blue!6,
            text width=4.0cm,
            align=left,
            font=\scriptsize,
            inner sep=4pt
        ] (proc) at (8.0,0.98)
        {2. Collect sensing snapshots $\Ybf$\\
         3. GLRT / Subspace-MUSIC Estimation,
         Output: $\hat{\mathcal K}$ and $\widehat{\wbf}$};

        % optional short connector to show processing at BS
        \draw[gray!70, thick, ->] (bs.south) -- (proc.north);

        %--------------------------------------------------
        % Desired NTN link
        %--------------------------------------------------、

        \draw[<->, thick, green!50!black] (sat) -- node[midway, above, sloped] {NTN-Link} (ntn);

        %--------------------------------------------------
        % Blue broadcast waves from NTN-UE
        %--------------------------------------------------
        \foreach \r in {0.42,0.68,0.94} {
            \draw[blue, very thick]
            ($(ntn.east)+(0.02,0.3)$) ++(-28:\r)
            % arc[start angle=-20, end angle=40, radius=\r];
            arc[start angle=-32, end angle=32, radius=\r];
        }
        \node[blue, align=center] at (4.5,3.62)
        {1. Uplink activity\\(unknown waveform)};

        %--------------------------------------------------
        % Red nulling arrow from TN-BS
        %--------------------------------------------------
        % \draw[->, thick, red]
        % (bs.west) to[bend left=12]
        % node[midway, below] {4. Noncoherent Nulling}
        % (ntn.east);
        \draw[->, thick, red]
        ($(bs.west)+(0,-0.05)$) to[bend left=12]
        node[midway, below=2pt, fill=white, inner sep=1pt] {4. Noncoherent Nulling}
        ($(ntn.east)+(0,-0.1)$);
        %--------------------------------------------------
        % Desired TN beam
        %--------------------------------------------------
        \draw[->, thick, green!50!black] (bs) --node[midway, below, sloped] {4. TN Beamforming} (tn);

        %--------------------------------------------------
        % Beam-design note
        %--------------------------------------------------
        \node[align=center] at (7.9,4.5)
        {Beam design from \\
        desired channel $\hbf_{0}$ and \\
        detected victim directions/gains};

    \end{tikzpicture}
\caption{Noncoherent sensing and interference nulling for terrestrial--satellite downlink coexistence. The TN-BS uses victim uplink activity to estimate the active victim set $\hat{\mathcal K}$ and direction/gain information, then forms a beam $\widehat{\wbf}$ that preserves the TN link while reducing leakage toward detected NTN users.}
    \label{fig:noncoh_framework}
    \vspace{-18pt}
\end{figure}

Prior work has studied TN-to-satellite coexistence interference~\cite{2024tractable} and developed mitigation methods including robust beamforming under channel uncertainty~\cite{lin2018robust}, interference nulling from TN base stations toward satellites~\cite{ULNulling, ULNulling2}, reverse spectrum sharing with joint TN/NTN grouping and beamforming~\cite{lee2024reverse}, and beamforming based on angular radiated power constraints~\cite{ueda2025angular}. More recent work has also considered in-band coexistence in dense LEO networks via satellite selection~\cite{kim2026satellite}. Existing mitigation methods generally assume known channel state information or coordinated signaling. In particular, ~\cite{jia2025joint},~\cite{NTN-Antijamming} relies on known preambles or beacons for victim detection and channel estimation. In contrast, the present method assumes that the victim waveform, pilot structure, and identity are unknown, and recovers only anonymous direction and gain information from the sample covariance for downlink nulling.

% These methods typically rely on structured prior information such as known geometry, channel state information, or coordinated signaling. In the downlink setting considered here, however, a terrestrial BS may have limited knowledge of the victim receivers and their transmissions, so pilot- or preamble-based sensing is not always available~\cite{jia2025joint}.

This paper studies \emph{noncoherent} victim sensing and interference nulling for this non-cooperative setting. The terrestrial BS listens to victim uplink activity, treats the temporal waveform as an unknown nuisance parameter, and extracts spatial information needed for interference suppression. It then combines the detected victim directions and powers with the desired TN channel to form a downlink beam that preserves the TN link while reducing leakage toward vulnerable NTN users.

\noindent \textbf{Our Contributions:}

\begin{itemize}
\item \emph{Noncoherent sensing and nulling framework}:  
We develop a unified noncoherent framework for victim sensing and interference nulling. The single-victim GLRT reduces to a principal-eigenvector estimator related to~\cite{GLRT}, and the multi-victim extension uses MDL-MUSIC to recover anonymous direction/gain tuples for downlink nulling.

% We develop a unified noncoherent framework for victim sensing and interference nulling. For a single victim, the GLRT reduces to a principal-eigenvector estimator of the sample covariance~\cite{GLRT}. For multiple simultaneous victims, we pair MDL-based source enumeration with MUSIC scanning to recover direction/gain tuples, and use them to design a transmit beam that preserves the desired TN link while suppressing interference toward detected victims.

\item \emph{Large-matrix detection error analysis}:  
Using the rank-one matrix estimation framework in~\cite{esterror}, we characterize the overlap between the GLRT estimate and the true victim direction. This yields the sensing-SNR scaling $1-\eta=O(\gamma_{\mathrm S}^{-1})$ and explains the residual leakage trend after nulling.

\item \emph{Site-specific coexistence evaluation}:  
We evaluate the proposed approach through site-specific ray-tracing simulations in a large TN--NTN deployment. The results show significant reduction in NTN interference with only modest degradation in TN link performance.

\end{itemize}

\section{Problem Formulation}

\paragraph*{Signal Model}
We consider a terrestrial transmitter (TX) with $N_{\mathrm{tx}}$ antennas serving an intended TN user $\RX{0}$ while coexisting with victim receivers $\RX{1},\ldots,\RX{N}$. Let $\wbf\in\C^{N_{\mathrm{tx}}\times 1}$ be the unit-norm transmit beamformer, $\bsym{h}_i\in\C^{N_{\mathrm{tx}}\times 1}$ the narrowband channel from the TX to $\RX{i}$, $\mc{E}_{\subsf D}$ the downlink symbol energy, and $N_0$ the receiver noise energy. Then the victim INR is
\begin{equation}\label{eq:inr}
    \INR_i = \frac{\mc{E}_{\subsf D}|\wbf^*\bsym{h}_i|^2}{N_0}, \quad i=1,\ldots,N.
\end{equation}
For $i=0$, the same expression gives the desired-link SNR.
\paragraph*{Victim Sensing Model}
During sensing, the TX collects $T$ array snapshots from victim uplink activity and models the received matrix as
\begin{equation}
\Ybf=\sum_{i=1}^{K}\ubf_i \bbf_i^{\T}+\Wbf,
\label{eq:sigmod}
\end{equation}
where $\Ybf\in\C^{N_{\mathrm{tx}}\times T}$, $K$ is the number of active victims, and each unit-norm vector $\ubf_i$ is the spatial signature of victim $i$. The temporal vector $\bbf_i=[b_i(1),\ldots,b_i(T)]^{\T}$ collects the $T$ uplink samples with $b_i(\tau)=g_is_i(\tau)$, so it absorbs both the effective sensing coefficient $g_i$ and the transmitted symbol $s_i(\tau)$. The noise matrix $\Wbf$ has i.i.d.\ entries distributed as $\CN(0,\sigma^2)$.

\section{Detection and Nulling}
\label{sec:detection_nulling}

\subsection{GLRT Estimator}
\label{sec:glrt_estimator}

For a single candidate victim, \eqref{eq:sigmod} reduces to the binary test
\begin{align}
H_0: \Ybf & = \Wbf, \nn\\
H_1: \Ybf & = \ubf_1\bbf^{\T}+\Wbf,\qquad \|\ubf_1\|_2=1.
\label{eq:hypos}
\end{align}
With circular Gaussian noise, the GLRT is
\begin{equation}
\Lambda(\Ybf)
=
\frac{\max_{\ubf_1,\bbf,\sigma^2} p(\Ybf\mid H_1,\ubf_1,\bbf,\sigma^2)}
{\max_{\sigma^2} p(\Ybf\mid H_0,\sigma^2)}
\hypcmp{H_1}{H_0}
\gamma_{\mathrm{G}},
\label{eq:glrt_ratio}
\end{equation}
where $\gamma_{\mathrm{G}}$ is chosen to meet the target false-alarm probability. 
For fixed unit-norm \(\mathbf u_1\), the unknown temporal waveform is a nuisance parameter. Maximizing over it is simply the least-squares projection of \(\mathbf Y\) onto the one-dimensional spatial subspace spanned by \(\mathbf u_1\), giving \(\widehat{\mathbf b}^{T}=\mathbf u_1^{H}\mathbf Y\). Thus the receiver need not decode the victim waveform; it only tests which spatial direction captures the largest received energy:
% For fixed unit-norm $\ubf_1$, profiling out the unknown waveform gives the least-squares estimate $\widehat{\bbf}^{\T}=\ubf_1^{\Her}\Ybf$, so the maximization under $H_1$ reduces to a Rayleigh-quotient problem in $\ubf_1$; Appendix~\ref{app:glrt_details} gives the explicit likelihoods and algebra.
\begin{equation}
\min_{\bbf}\|\Ybf-\ubf_1\bbf^{\T}\|_F^2
=
\|\Ybf\|_F^2-\|\ubf_1^{\Her}\Ybf\|_2^2,
\label{eq:glrt_profile_cost}
\end{equation}
This makes the Rayleigh-quotient structure explicit, and the resulting spatial estimate is the principal eigenvector of the sample covariance
\begin{equation}
\Rbf
\triangleq
\frac{1}{T}\Ybf\Ybf^{\Her},
\qquad
\widehat{\ubf}_1
=
\vbf_{\max}(\Rbf).
\label{eq:glrt_uhat}
\end{equation}
Under $H_0$, the ML noise-variance estimate is
\begin{equation}
\widehat{\sigma}_0^2
=
\frac{\|\Ybf\|_F^2}{N_{\mathrm{tx}}T},
\label{eq:glrt_sigma0}
\end{equation}
while after profiling out $\bbf$ and $\ubf_1$ under $H_1$,
\begin{equation}
\widehat{\sigma}_1^2
=
\frac{\|\Ybf\|_F^2-\lambda_{\max}(\Ybf\Ybf^{\Her})}{N_{\mathrm{tx}}T}.
\label{eq:glrt_sigma1}
\end{equation}
Hence the profiled GLRT statistic becomes
\begin{equation}
\Lambda(\Ybf)
=
\left(
\frac{\widehat{\sigma}_0^2}{\widehat{\sigma}_1^2}
\right)^{N_{\mathrm{tx}}T}.
\label{eq:glrt_profiled}
\end{equation}
The direction estimate is analyzed in Section~\ref{sec:large_matrix}, and the profiled GLRT is monotone in the normalized largest eigenvalue
% This direction estimate is the quantity analyzed in Section~\ref{sec:large_matrix}. The profiled GLRT is monotone in the normalized largest eigenvalue
\begin{equation}
\xi
\triangleq
\frac{\lambda_{\max}(\Rbf)}{\Tr(\Rbf)},
\label{eq:glrt_xi}
\end{equation}
and can therefore be implemented as
\begin{equation}
\lambda_{\max}(\Rbf)
\hypcmp{H_1}{H_0}
\psi\,\Tr(\Rbf),
\label{eq:glrt_detector}
\end{equation}
for an appropriate threshold $\psi$. Thus, the GLRT provides both the single-candidate detection rule and the spatial estimate $\widehat{\ubf}_1$ used later in the large-matrix analysis.
\subsection{Blind MDL-MUSIC Detection}
\label{sec:subspace_music}
For multiple simultaneous victims, the general model \eqref{eq:sigmod} applies with unknown source count $K$, unit-norm steering vectors $\ubf_i=\ubf(\phi_i,\theta_i)$, and noncoherent source powers $G_i\triangleq T^{-1}\|\bbf_i\|_2^2$. If the empirical cross-correlations $\frac{1}{T}\bbf_i^{\Her}\bbf_j$ are negligible for $i\neq j$, then the sample covariance obeys
\begin{equation}
\widehat{\Rbf}
=
\frac{1}{T}\Ybf\Ybf^{\Her}
\approx
\sum_{i=1}^{K} G_i \ubf(\phi_i,\theta_i)\ubf(\phi_i,\theta_i)^{\Her}
\,+\,\sigma^2\Ibf .
\label{eq:music_cov}
\end{equation}
This approximation requires sufficiently many snapshots and weak empirical source correlation. Correlated or synchronized uplink transmissions, angularly close victims, severe near-far imbalance, or array-calibration errors can bias the sample covariance and degrade both MDL model-order selection and MUSIC peak finding. Within this regime, the sensing stage remains blind: it does not decode pilots or identify users, but only recovers anonymous direction/power tuples for beam design.
% Thus, the sensing stage is blind: it does not decode pilots or identify users, but only recovers anonymous direction/power tuples for beam design.

Let $\hat{\lambda}_1\ge\cdots\ge\hat{\lambda}_{N_{\mathrm{tx}}}$ denote the ordered eigenvalues of $\widehat{\Rbf}$. The active-source count is selected by the Wax--Kailath MDL rule
\begin{equation}
\hat K
=
\argmin_{0\leq k\leq N_{\mathrm{tx}}-1}\mathrm{MDL}(k),
\label{eq:music_mdl_k}
\end{equation}
where the explicit criterion is given in Appendix~\ref{app:music_details} by \eqref{eq:appB_mdl_def}. The corresponding noise subspace $\Ubf_{\mathrm n}$ is formed from the trailing eigenvectors $\hat{\lambda}_{\hat K+1},\ldots,\hat{\lambda}_{N_{\mathrm{tx}}}$. We therefore refer to this stage as \emph{Blind MDL-MUSIC Detection}. It then searches for dominant directions:
\begin{equation}
\mc{P}_{\mathrm{MU}}(\phi,\theta)
=
\frac{1}{
\ubf(\phi,\theta)^{\Her}\Ubf_{\mathrm{n}}\Ubf_{\mathrm{n}}^{\Her}\ubf(\phi,\theta)
+\epsilon_{\mathrm{M}}
},
\label{eq:music_detect}
\end{equation}
where $\epsilon_{\mathrm{M}}>0$ is a small regularization constant. The $\hat K$ dominant peaks of \eqref{eq:music_detect} over the angular grid define the estimated directions $\{(\hat{\phi}_{k},\hat{\theta}_{k})\}_{k=1}^{\hat K}$, and the associated gains are obtained by the covariance-domain nonnegative least-squares fit
\begin{equation}
\{\hat{G}_{k}\}_{k=1}^{\hat K}
=
\argmin_{\{G_k\ge 0\}}
\left\|
\widehat{\Rbf}
-\hat{\sigma}^{2}\Ibf
-\sum_{k=1}^{\hat K} G_k \hat{\ubf}_{k}\hat{\ubf}_{k}^{\Her}
\right\|_{F}^{2},
\label{eq:music_gain_est}
\end{equation}
where $\hat{\ubf}_{k}=\ubf(\hat{\phi}_{k},\hat{\theta}_{k})$ and $\hat{\sigma}^{2}$ is estimated from the smallest eigenvalues of $\widehat{\Rbf}$. The detector therefore outputs the anonymous tuple set $\{(\hat{\phi}_{k},\hat{\theta}_{k},\hat{G}_{k})\}_{k=1}^{\hat K}$ used by the nulling stage.

\subsection{Noncoherent Nulling}
\label{sec:noncoherent_nulling}
Once the estimated interference directions and gains are available, the TX forms its transmit beam from the desired effective channel $\hbf_{0}$ and the detected interference subspace. Here, $\hbf_{0}$ denotes the effective channel of the intended TN user. The beamformer is obtained from
\begin{equation}
\max_{\|\wbf\|_2=1}
\;
|\wbf^{\Her}\hbf_{0}|^2
-\lambda\sum_{k=1}^{\hat K}
\hat{G}_{k}
\big|
\wbf^{\Her}\ubf(\hat{\phi}_{k},\hat{\theta}_{k})
\big|^2.
\label{eq:noncoh_obj}
\end{equation}
Equivalently, define the Hermitian design matrix
\begin{equation}
\Qbf
=
\hbf_{0}\hbf_{0}^{\Her}
-\lambda\sum_{k=1}^{\hat K}
\hat{G}_{k}\,
\ubf(\hat{\phi}_{k},\hat{\theta}_{k})
\ubf(\hat{\phi}_{k},\hat{\theta}_{k})^{\Her}.
\label{eq:noncoh_Q}
\end{equation}
This construction weights stronger detected interference directions more heavily. The parameter \(\lambda\) sets the TN/NTN tradeoff and can be chosen from an INR target or TN-SINR constraint: increasing \(\lambda\) enforces stricter nulling, while decreasing it favors the desired link. Complexity is dominated by covariance eigendecomposition, MUSIC grid search, and the final eigenvector computation, yielding a latency tradeoff with grid resolution and sensing-update rate. The optimal beamformer is
% This construction places larger penalties on stronger interference directions through the weights $\hat{G}_{k}$. The optimal beamformer is therefore
\begin{equation}
\widehat{\wbf}
=
\vbf_{\max}(\Qbf).
\label{eq:noncoh_sol}
\end{equation}

\section{Large-Matrix Detection Error Rates}
\label{sec:large_matrix}

This section studies the accuracy of the GLRT spatial estimate in \eqref{eq:glrt_uhat} for the single-victim specialization of \eqref{eq:sigmod} with $K=1$. Consistent with the signal model in Section~II, we write the unknown temporal factor as $\bbf=g\sbf$, where $g$ is the effective uplink sensing coefficient, $\sbf\in\C^T$ collects the victim uplink symbols, and the entries of $\Wbf$ are i.i.d.\ $\CN(0,\sigma^2)$. We define the sensing SNR as
\begin{equation}
\gamma_{\mathrm{S}}
\triangleq
\frac{|g|^2}{\sigma^2}.
\label{eq:lm_gamma_s}
\end{equation}

We quantify the estimation quality through the overlap
\begin{equation}
\eta
\triangleq
\big|\widehat{\ubf}_1^{\Her}\ubf_1\big|^2.
\label{eq:lm_eta_def}
\end{equation}
This overlap can be precisely quantified in the large system limit
using results from \cite{esterror}.
Specifically, suppose $N_{\mathrm{tx}},T\rightarrow\infty$ and fixed aspect ratio
\begin{equation}
\beta
\triangleq
\frac{T}{N_{\mathrm{tx}}},
\label{eq:lm_beta}
\end{equation}
the model matches the rank-one model in \cite{esterror} with 
\begin{equation}
\begin{aligned}
\tau_u
&=
\gamma_{\mathrm{S}}\Exp\!\left[|u_{1,m}|^2\right],
&
\tau_v
&=
\Exp\!\left[|s(\ell)|^2\right],\\
\tau_w
&=
\frac{\Exp\!\left[|W_{m,\ell}|^2\right]}{N_{\mathrm{tx}}},
&
\alpha
&\triangleq
\frac{\tau_w}{\tau_u\tau_v}.
\end{aligned}
\label{eq:lm_tau_map}
\end{equation}
Here $u_{1,m}$ denotes the $m$th entry of the victim spatial signature $\ubf_1$, $s(\ell)$ is the $\ell$th uplink symbol in $\sbf$, and $W_{m,\ell}$ is the corresponding noise sample. Under the unit-power signaling assumption adopted here, $\tau_v=1$.
Then from results in \cite{esterror}, the overlap obeys
\begin{equation}
\eta
\approx
\frac{\left[\beta-\alpha^2\right]_+}{\beta+\alpha}
=
1-\frac{\alpha}{\beta}+O(\alpha^2).
\label{eq:lm_eta}
\end{equation}
For fixed $\beta$ and unit-power uplink symbols, \eqref{eq:lm_tau_map} implies $\alpha=O(\gamma_{\mathrm{S}}^{-1})$, and therefore
\begin{equation}
1-\eta
=
O(\gamma_{\mathrm{S}}^{-1}).
\label{eq:lm_eta_rate}
\end{equation}
Thus, a stronger victim uplink drives the GLRT eigenvector toward the true victim direction. The validation below compares this overlap trend and the resulting nulling leakage against the retained-$\delta$ model derived in Appendix~\ref{app:large_matrix_interpretation}.
\begin{figure}[!t]
\centering
\includegraphics[width=0.65\columnwidth]{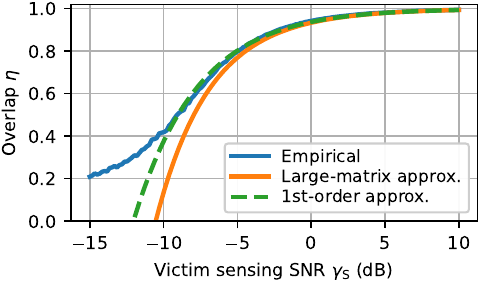}
\caption{GLRT overlap validation for a one-target/one-victim ULA setting with target angle $\theta_0=0^\circ$, victim angle $\theta_1=18^\circ$, $N_{\mathrm{tx}}=8$, and $T=16$. The two model curves correspond to the large-matrix overlap approximation and its first-order expansion in \eqref{eq:lm_eta}.}
\label{fig:lm_eta_validation}
\vspace{-12pt}
\end{figure}

\begin{figure}[!t]
\centering
\includegraphics[width=0.65\columnwidth]{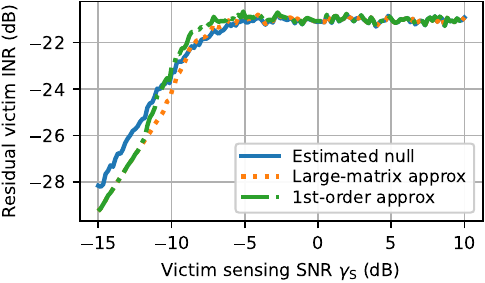}
\caption{Residual victim INR versus victim uplink sensing SNR. The ``Estimated null'' curve uses the single-target/single-victim specialization of the Section~\ref{sec:noncoherent_nulling} beamformer. The retained-$\delta$ model curves evaluate \eqref{eq:lm_inr_exact}, using the empirical overlap, the large-matrix overlap approximation in \eqref{eq:lm_eta}, and its first-order expansion $1-\alpha/\beta$.}
\label{fig:lm_inr_validation}
\vspace{-15pt}
\end{figure}

Figure~\ref{fig:lm_eta_validation} confirms the overlap analysis: both the large-matrix approximation in \eqref{eq:lm_eta} and its first-order expansion track the empirical GLRT overlap and its monotonic improvement with sensing SNR. To connect this overlap error to post-nulling leakage, let $\ubf_0$ denote the target direction and write $\widehat{\ubf}_1=\sqrt{\eta}\,\ubf_1+\ebf$, where $\ebf$ is the residual estimation error orthogonal to $\ubf_1$ so that $\ubf_1^{\Her}\ebf=0$ and $\|\ebf\|_2^2=1-\eta$. Define $\rho\triangleq \ubf_1^{\Her}\ubf_0$ and $\delta\triangleq \ebf^{\Her}\ubf_0$; then the effective target projection becomes $\hat{\rho}\triangleq \widehat{\ubf}_1^{\Her}\ubf_0=\sqrt{\eta}\,\rho+\delta$. Let $\gamma_{\mathrm{I}}$ denote the pre-nulling victim INR scale in the downlink leakage expression, which is distinct from the uplink sensing SNR $\gamma_{\mathrm{S}}$. The retained-$\delta$ residual victim INR is
\begin{equation}
\INR
=
\gamma_{\mathrm{I}}
\frac{\left|(1-\eta)\rho-\sqrt{\eta}\,\delta^*\right|^2}
{1-\left|\sqrt{\eta}\,\rho+\delta\right|^2}.
\label{eq:lm_inr_exact}
\end{equation}
Figure~\ref{fig:lm_inr_validation} shows how the overlap error translates into leakage. Under the coupled-geometry interpretation, the same pathloss trend that increases $\gamma_{\mathrm{S}}$ also increases $\gamma_{\mathrm{I}}$. As the victim uplink sensing SNR first increases, the post-nulling INR rises because the victim is effectively closer to the base station and the raw interference is stronger. Once the victim becomes sufficiently strong, the INR saturates because the GLRT spatial estimate improves enough to compensate for the smaller path loss. Substituting the two overlap approximations into \eqref{eq:lm_inr_exact} reproduces this trend and closely matches the estimated-null curve; details are provided in Appendix~\ref{app:large_matrix_interpretation}.

% Figure~\ref{fig:lm_inr_validation} shows how this overlap error translates into leakage. In the coupled-geometry interpretation used here, the same pathloss trend that increases $\gamma_{\mathrm{S}}$ also increases $\gamma_{\mathrm{I}}$, as detailed in Appendix~\ref{app:large_matrix_interpretation}. As the victim uplink sensing SNR first increases, the post-nulling INR rises because the victim is effectively closer to the base station and the raw interference is stronger. Once the victim becomes strong, the INR saturates because the GLRT spatial estimate improves enough to compensate for the reduced path loss. Substituting the two overlap approximations into \eqref{eq:lm_inr_exact} reproduces this trend and stays close to the estimated-null curve; Appendix~\ref{app:large_matrix_interpretation} provides the full derivation.

\section{Simulation Results and Performance Evaluation}

\subsection{Multi-Victim Scenario Evaluation}

\begin{figure}[!t]
    \centering
    \includegraphics[width=0.35\textwidth]{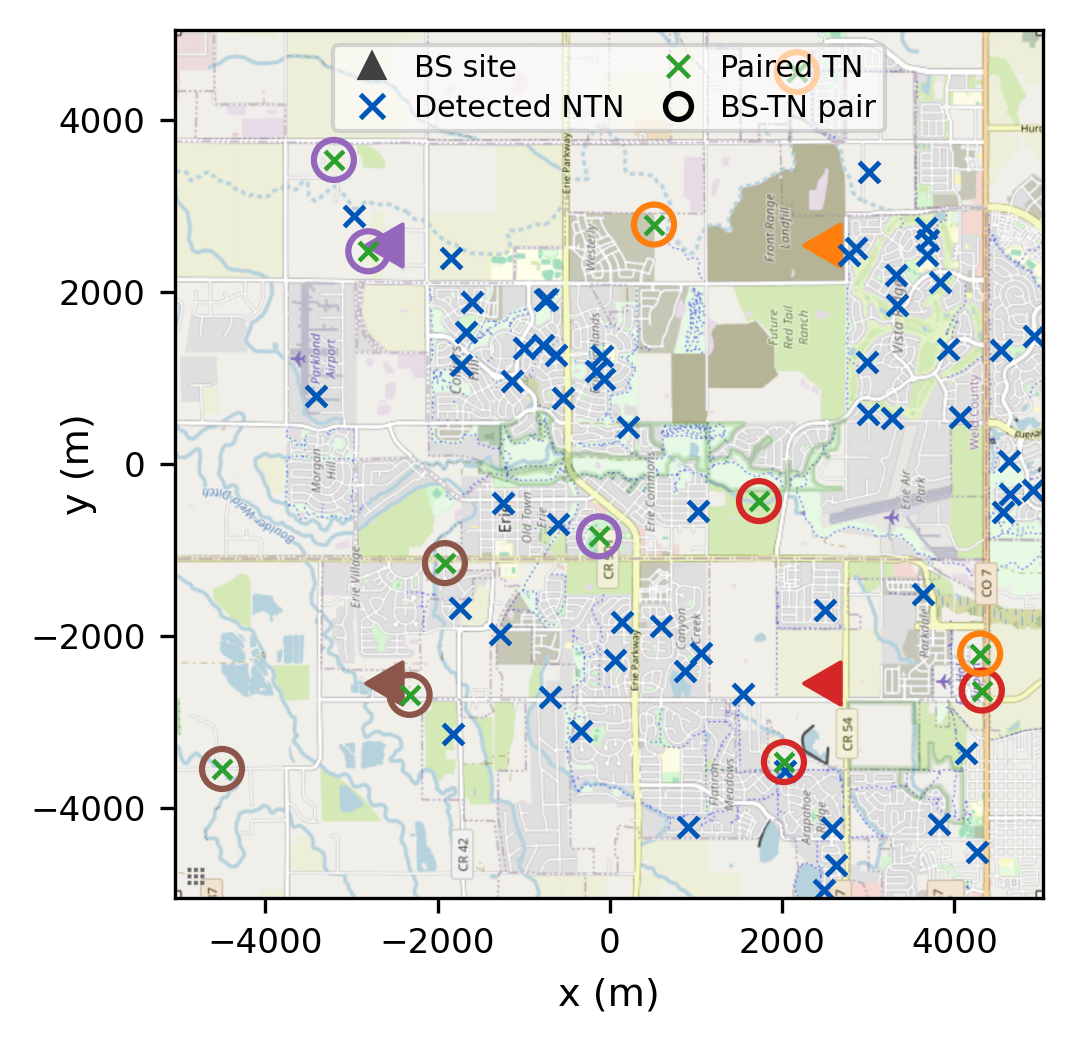}
    \vspace{-5pt}
    \caption{Representative site-specific snapshot for the multi-victim nulling evaluation. Triangles mark the four TN-BS sites, blue crosses denote detected NTN terminals, and green crosses denote the TN users scheduled in the current round. Each circled TN marker indicates an active BS--TN pair, with the circle color identifying the serving BS. }
    \label{fig:map}
    \vspace{-20pt}
\end{figure}

We evaluate the proposed scheme through Monte Carlo simulations over a \SI{10}{\kilo\meter}$\times$\SI{10}{\kilo\meter} rural area near Boulder, Colorado, shown in Fig.~\ref{fig:map}, using the Sionna ray-tracing platform~\cite{sionna}. The terrestrial network consists of four outdoor BS sites on a $2\times2$ grid, each with three sectors, while 300 TN-UEs and 100 NTN-UEs (VSATs) are randomly deployed in each macro realization. The main carrier, antenna, deployment, and propagation parameters are summarized in Table~\ref{tab:simulation_parameters}. To capture satellite-geometry variability, the common NTN look direction is resampled in each realization from a \SI{550}{\kilo\meter} LEO orbit, with azimuth uniformly distributed over $[0^\circ,360^\circ]$ and elevation over $[45^\circ,90^\circ]$. We generate 100 macro realizations. In each realization, each TN-UE is associated with the strongest sector meeting the TN-link threshold, and each active sector serves one paired TN-UE per round. Victim sensing follows the subspace-MUSIC procedure in Section~\ref{sec:subspace_music}, and performance is reported by the CDFs of NTN INR and TN SNR/SINR.

\begin{table}[!t]
\centering
\caption{\textbf{Simulation Parameters.}}
\label{tab:simulation_parameters}
\renewcommand{\arraystretch}{1.3}
\large
\resizebox{0.48\textwidth}{!}{
\begin{tabular}{|l|c|}
\hline
\textbf{Radio Parameters} & \textbf{Values} \\ \hline
Carrier Frequency and Bandwidth & \SI{10}{\giga\hertz} and \SI{200}{\mega\hertz} \\ \hline
Noise Power Spectral Density & \SI{-174}{dBm\per\hertz} \\ \hline
Transmit Power & \SI{40}{dBm} (BS), \SI{35}{dBm} (VSAT), \SI{23}{dBm} (Handheld)\\ \hline
Noise Figure & \SI{3}{\dB} (TN-BS), \SI{2}{\dB} (VSAT), \SI{7}{\dB} (Handheld) \\ \hline
Multipath order & 3 (Reflection and Penetration) \\ \hline
Base Station (BS) & $8 \times 8$ URA, Directional~\cite{3gpp.38.901}, 3 sectors \\ \hline
Handheld (TN-UE) & (1, 1, 2) Omni-directional~\cite{3gpp.38.821} \\ \hline
VSAT (NTN-UE) & \SI{60}{\centi\meter} diameter, Directional~\cite{3gpp.38.821} \\ \hline
Antenna Heights & \SI{40}{\meter} (TN-BS), \SI{1.6}{\meter} (outdoor), \SI{1}{\meter} (rooftop) \\ \hline

\multicolumn{2}{|l|}{\textbf{Scene Configuration}} \\ \hline
Geographical Location  & \SI{10}{\kilo\meter} $\times$ \SI{10}{\kilo\meter} rural area \\ \hline
Radio Material & Wall: Brick, Roof: Concrete, Ground: Medium Dry \\ \hline
Satellite Angle & Azimuth: [0°, 360°], Elevation: [45°, 90°] \\ \hline
Satellite Orbit & \SI{550}{\kilo\meter} (LEO) \\ \hline
NTN-UE Deployment & 80 rooftop, 20 outdoor\\ \hline
TN-UE Deployment & 180 indoor, 120 outdoor \\ \hline
TN-BS Deployment & $2\times 2$ layout, \SI{5}{\degree} downtilt \\ \hline
\end{tabular}}
\vspace{-10pt}
\end{table}

\subsection{Terrestrial Downlink Interference Nulling}
\begin{figure}[!t]
\centering
\includegraphics[width=0.36\textwidth]{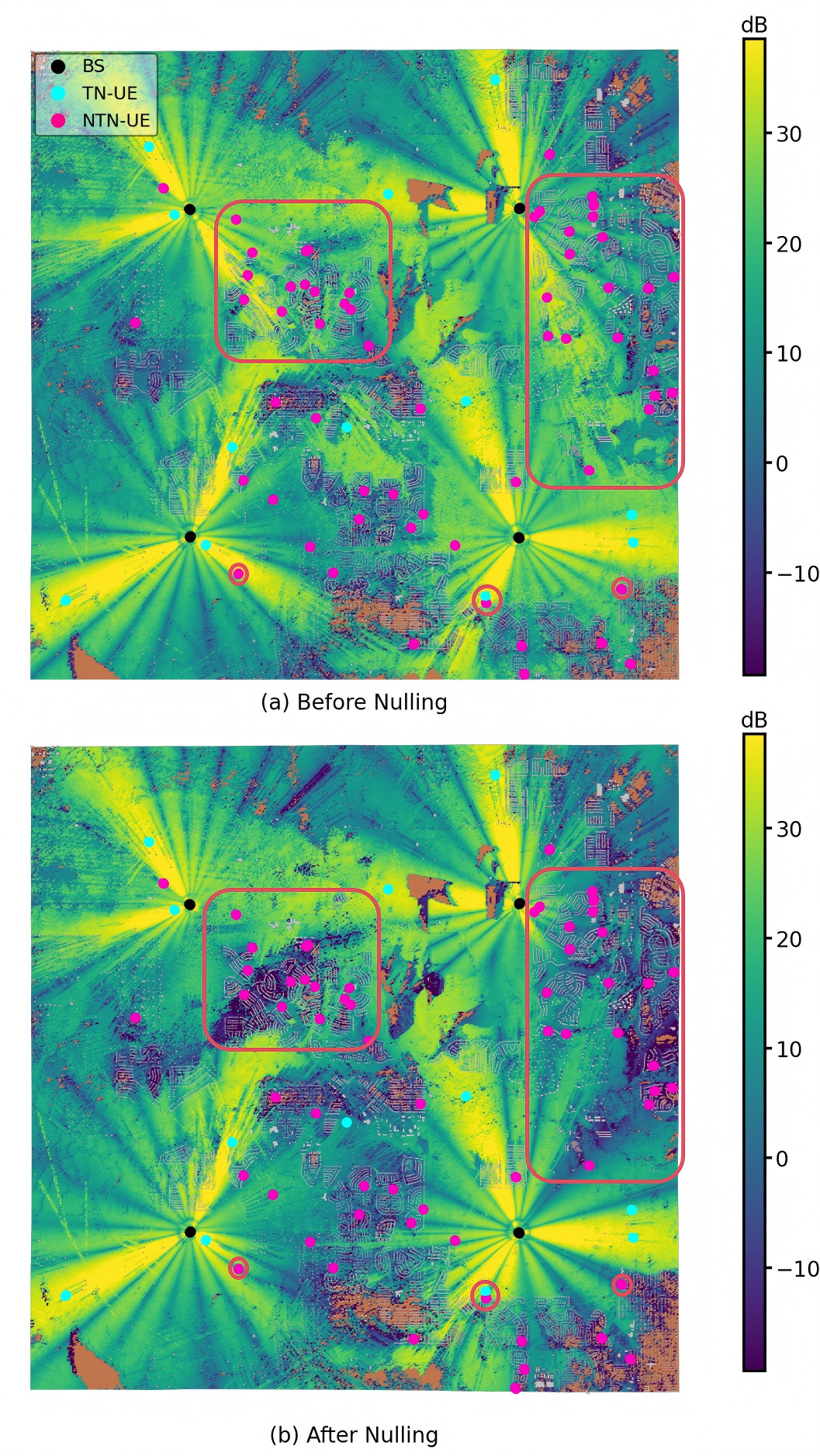}
\vspace{-5pt}
\caption{INR Radio map before and after applying the proposed nulling method. Each base station sector forms spatial nulls toward detected NTN-UE directions, with pronounced interference suppression for NTN-UEs highlighted especially in the red-circled region.(Same ditribution with Fig. \ref{fig:map})}
\label{fig:radiomap}
\vspace{-15pt}
\end{figure}
\begin{figure}[!t]
\centering
\includegraphics[width=0.36\textwidth]{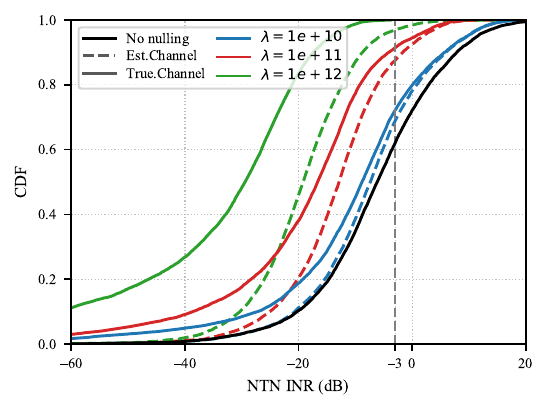}
\vspace{-3pt}
\caption{Empirical CDF of the downlink INR at the detected NTN victims. The black curve is the no-nulling baseline, dashed colored curves correspond to MUSIC-detection noncoherent nulling with estimated victim directions and gains, while solid colored curves show the same design using the true detected-user information for different values of~$\lambda$.}
\label{fig:inr_cdf}
\vspace{-18pt}
\end{figure}
\begin{figure}[!t]
\centering
\includegraphics[width=0.36\textwidth]{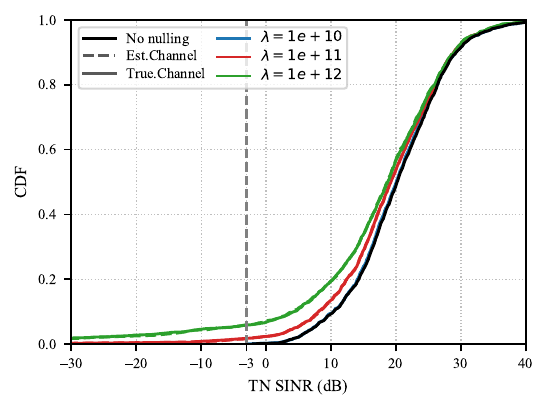}
\vspace{-3pt}
\caption{Empirical CDF of the TN-users SINR under the same regularization sweep as in Fig.~\ref{fig:inr_cdf}. The legend follows the same convention.}
\label{fig:sinr_cdf}
\vspace{-15pt}
\end{figure}

An NTN terminal may receive interference from multiple simultaneously active BS sectors. Let $\mathcal{B}_{\mathrm{act}}$ denote the set of active sectors in one scheduling round. If sector $b\in\mathcal{B}_{\mathrm{act}}$ uses beamformer $\widehat{\wbf}_b$ and the channel from sector $b$ to NTN terminal $i$ is $\bsym{h}_{i,b}$, then the aggregate downlink INR reported in Figs.~\ref{fig:radiomap}--\ref{fig:sinr_cdf} is
\begin{equation}
\INR_i^{\mathrm{net}}
=
\frac{1}{N_0}
\sum_{b\in\mathcal{B}_{\mathrm{act}}}
\mc{E}_{\subsf D,b}
\left|
\widehat{\wbf}_b^{\Her}\bsym{h}_{i,b}
\right|^2 .
\label{eq:network_inr}
\end{equation}
Fig.~\ref{fig:radiomap} shows the aggregate INR field under the same site-specific deployment for $\lambda=10^{12}$. Compared with the no-nulling basemap, the proposed nulling suppresses the high-INR regions around the detected NTN terminals.

\paragraph*{NTN INR}\label{INR}
Fig.~\ref{fig:inr_cdf} shows that the nulling shifts the INR CDF of the detected NTN terminals to the left as $\lambda$ increases. Relative to the no-nulling baseline, the fraction of samples below \SI{-3}{dB} increases from $62\%$ to $69\%$, $88\%$, and $97\%$ under the estimated-channel design for $\lambda=10^{10},10^{11},10^{12}$, respectively. With the true-channel benchmark, the corresponding fractions further improve to $72\%$, $92\%$, and $100\%$. This confirms that larger $\lambda$ provides stronger NTN protection, while the gap between the two designs shows that channel estimation error limits residual leakage at large $\lambda$.

\paragraph*{TN SINR}\label{SINR}
Fig.~\ref{fig:sinr_cdf} shows that the TN-side penalty remains modest. For $\lambda=10^{10}$, the SINR CDF is almost unchanged from the no-nulling baseline. As $\lambda$ increases to $10^{11}$ and $10^{12}$, the median TN SINR decreases only from about \SI{20.2}{dB} to \SI{19.2}{dB} and \SI{18.7}{dB}, respectively. Moreover, the estimated-channel and true-channel curves are nearly overlapping, indicating that channel estimation error has a much smaller impact on TN performance than on NTN leakage suppression.

% \paragraph*{NTN INR}\label{INR}
% Fig.~\ref{fig:inr_cdf} shows that the proposed nulling shifts the INR CDF of the detected NTN terminals to the left as $\lambda$ increases. Relative to the no-nulling baseline, the fraction of samples below \SI{-3}{dB} increases from $62\%$ to $69\%$, $88\%$, and $97\%$ under the estimated-channel design for $\lambda=10^{10},10^{11},10^{12}$, respectively. With the true-channel benchmark, the corresponding fractions further improve to $72\%$, $92\%$, and $100\%$. This confirms that larger $\lambda$ provides stronger NTN protection, while the gap between the two designs shows that channel estimation error mainly limits the residual leakage at large $\lambda$.

% \paragraph*{TN SINR}\label{SINR}
% Fig.~\ref{fig:sinr_cdf} shows that the TN-side penalty remains modest. For $\lambda=10^{10}$, the SINR CDF is almost unchanged from the no-nulling baseline. As $\lambda$ increases to $10^{11}$ and $10^{12}$, the median TN SINR decreases only from about \SI{20.2}{dB} to \SI{19.2}{dB} and \SI{18.7}{dB}, respectively. Moreover, the estimated-channel and true-channel curves are nearly overlapping, indicating that channel estimation error has a much smaller impact on TN performance than on NTN leakage suppression.
\label{results}

\section{Conclusion}
\label{conclusion}

This paper presented a noncoherent sensing and nulling framework for terrestrial--satellite downlink coexistence in the upper mid-band. Treating the victim uplink waveform as unknown, the single-victim GLRT reduces to a covariance-eigenvector estimator, while a Blind MDL-MUSIC front end extends the framework to multiple simultaneous victims and returns the anonymous direction/gain tuples needed for beam design. We further showed in the large-matrix regime that the GLRT overlap improves with sensing SNR, explaining the resulting leakage trend after nulling. Site-specific ray-tracing results showed substantial NTN INR reduction with only modest TN SINR loss.

\bibliographystyle{IEEEtran}
\bibliography{references}
\clearpage
\appendices
\section{Profile Likelihood Derivation for the GLRT}
\label{app:glrt_details}

This appendix records the algebra behind the profiled GLRT in Section~\ref{sec:glrt_estimator}. Under the two hypotheses in \eqref{eq:hypos}, the Gaussian likelihoods are
\begin{align}
p(\Ybf\mid H_0,\sigma^2)
&=
\frac{1}{(\pi\sigma^2)^{N_{\mathrm{tx}}T}}
\exp\!\left(
-\frac{\|\Ybf\|_F^2}{\sigma^2}
\right),
\nn\\
p(\Ybf\mid H_1,\ubf_1,\bbf,\sigma^2)
&=
\frac{1}{(\pi\sigma^2)^{N_{\mathrm{tx}}T}}
\exp\!\left(
-\frac{\|\Ybf-\ubf_1\bbf^{\T}\|_F^2}{\sigma^2}
\right).
\label{eq:appA_likelihoods}
\end{align}
Maximizing the likelihood under $H_0$ gives
\begin{equation}
\widehat{\sigma}_0^2
=
\frac{\|\Ybf\|_F^2}{N_{\mathrm{tx}}T}.
\label{eq:appA_sigma0}
\end{equation}
For fixed unit-norm $\ubf_1$, maximizing the likelihood under $H_1$ is equivalent to minimizing
\begin{equation}
J(\bbf)
\triangleq
\|\Ybf-\ubf_1\bbf^{\T}\|_F^2.
\label{eq:appA_cost}
\end{equation}
Expanding the quadratic form gives
\begin{align}
J(\bbf)
&=
\Tr\!\left\{(\Ybf-\ubf_1\bbf^{\T})(\Ybf-\ubf_1\bbf^{\T})^{\Her}\right\}
\nn\\
&=
\|\Ybf\|_F^2
-2\Re\!\left\{\bbf^{\T}\Ybf^{\Her}\ubf_1\right\}
+(\ubf_1^{\Her}\ubf_1)\bbf^{\Her}\bbf.
\label{eq:appA_expand}
\end{align}
Taking the derivative with respect to $\bbf^*$ and using $\|\ubf_1\|_2=1$ yields
\begin{equation}
\widehat{\bbf}
=
\Ybf^{\Her}\ubf_1,
\qquad
\widehat{\bbf}^{\T}
=
\ubf_1^{\Her}\Ybf,
\label{eq:appA_bhat}
\end{equation}
which is the least-squares estimate used to profile out the unknown waveform. Substituting \eqref{eq:appA_bhat} into \eqref{eq:appA_cost} gives
\begin{align}
\|\Ybf-\ubf_1\widehat{\bbf}^{\T}\|_F^2
&=
\|\Ybf-\ubf_1\ubf_1^{\Her}\Ybf\|_F^2
\nn\\
&=
\|\Ybf\|_F^2-\|\ubf_1^{\Her}\Ybf\|_2^2,
\label{eq:appA_profile}
\end{align}
Therefore, maximizing the profiled likelihood under $H_1$ reduces to a Rayleigh-quotient problem in $\ubf_1$, so the ML spatial estimate is the dominant eigenvector in \eqref{eq:glrt_uhat}.

With $\widehat{\ubf}_1$ and $\widehat{\bbf}$ substituted back into the likelihood, the ML estimate of the noise variance under $H_1$ is
\begin{equation}
\widehat{\sigma}_1^2
=
\frac{\|\Ybf\|_F^2-\lambda_{\max}(\Ybf\Ybf^{\Her})}{N_{\mathrm{tx}}T}.
\label{eq:appA_sigma1}
\end{equation}
Hence the profiled likelihood ratio becomes
\begin{equation}
\Lambda(\Ybf)
=
\left(
\frac{\widehat{\sigma}_0^2}{\widehat{\sigma}_1^2}
\right)^{N_{\mathrm{tx}}T}
=
\left(
\frac{\|\Ybf\|_F^2}{\|\Ybf\|_F^2-\lambda_{\max}(\Ybf\Ybf^{\Her})}
\right)^{N_{\mathrm{tx}}T}.
\label{eq:appA_glrt_ratio}
\end{equation}
Since $\|\Ybf\|_F^2=T\Tr(\Rbf)$ and $\lambda_{\max}(\Ybf\Ybf^{\Her})=T\lambda_{\max}(\Rbf)$, \eqref{eq:appA_glrt_ratio} is monotone in the normalized largest eigenvalue $\xi$ defined in \eqref{eq:glrt_xi}, so thresholding $\xi$ implements the GLRT.

\section{Blind MDL-MUSIC Detection Justification}
\label{app:music_details}

This appendix summarizes the modeling steps behind the blind detector in \eqref{eq:music_detect}. For sector $t$, the received snapshot matrix follows the multi-source model
\begin{equation}
\Ybf_t
=
\sum_{k=1}^{K_t}\ubf(\phi_{k,t},\theta_{k,t})\bbf_{k,t}^{\T}
+\Wbf_t,
\label{eq:appB_response_model}
\end{equation}
where each steering vector is unit norm, the temporal vectors $\{\bbf_{k,t}\}$ are unknown, and no known pilot structure is assumed. Define the noncoherent source powers
\begin{equation}
G_{k,t}
\triangleq
\frac{1}{T}\|\bbf_{k,t}\|_2^2 .
\label{eq:appB_gain_def}
\end{equation}
If the empirical cross-terms $\frac{1}{T}\bbf_{i,t}^{\Her}\bbf_{j,t}$ are negligible for $i\neq j$, then the sample covariance admits the approximation
\begin{equation}
\widehat{\Rbf}_t
\triangleq
\frac{1}{T}\Ybf_t\Ybf_t^{\Her}
\approx
\sum_{k=1}^{K_t} G_{k,t}\ubf(\phi_{k,t},\theta_{k,t})\ubf(\phi_{k,t},\theta_{k,t})^{\Her}
\,+\,\sigma_t^2\Ibf ,
\label{eq:appB_cov_model}
\end{equation}
which is the anonymous covariance model used by the detector in Section~\ref{sec:subspace_music}.

Let the ordered eigenvalues of $\widehat{\Rbf}_t$ be
\begin{equation}
\hat{\lambda}_{1,t}\geq \hat{\lambda}_{2,t}\geq\cdots\geq \hat{\lambda}_{N_{\mathrm{tx}},t},
\label{eq:appB_eigs}
\end{equation}
with associated eigenvectors $\{\qbf_{m,t}\}_{m=1}^{N_{\mathrm{tx}}}$. The model order is then estimated by the Wax--Kailath MDL rule
\begin{equation}
\hat K_t
=
\argmin_{0\leq k\leq N_{\mathrm{tx}}-1}\mathrm{MDL}_t(k),
\label{eq:appB_mdl_k}
\end{equation}
where
\begin{equation}
\begin{aligned}
\mathrm{MDL}_t(k)
&=
-T\big(N_{\mathrm{tx}}-k\big)
\log\!\left(
\frac{
\left(
\prod_{m=k+1}^{N_{\mathrm{tx}}}\hat{\lambda}_{m,t}
\right)^{\frac{1}{N_{\mathrm{tx}}-k}}
}{
\frac{1}{N_{\mathrm{tx}}-k}
\sum_{m=k+1}^{N_{\mathrm{tx}}}\hat{\lambda}_{m,t}
}
\right)
\\
&\quad
+\frac{1}{2}k\big(2N_{\mathrm{tx}}-k\big)\log T.
\end{aligned}
\label{eq:appB_mdl_def}
\end{equation}
From $\widehat{\Rbf}_t$, the MDL step estimates the signal-subspace dimension $\hat K_t$ and yields the decomposition
\begin{equation}
\widehat{\Rbf}_t
\approx
\Ubf_{\mathrm{s},t}\mathbf{\Lambda}_{\mathrm{s},t}\Ubf_{\mathrm{s},t}^{\Her}
+\Ubf_{\mathrm{n},t}\mathbf{\Lambda}_{\mathrm{n},t}\Ubf_{\mathrm{n},t}^{\Her},
\label{eq:appB_subspace_decomp}
\end{equation}
where
\begin{equation}
\Ubf_{\mathrm{s},t}
=
\left[\qbf_{1,t},\ldots,\qbf_{\hat K_t,t}\right],
\qquad
\Ubf_{\mathrm{n},t}
=
\left[\qbf_{\hat K_t+1,t},\ldots,\qbf_{N_{\mathrm{tx}},t}\right].
\label{eq:appB_subspaces}
\end{equation}
For a candidate steering direction $\ubf(\phi,\theta)$, the quantity
\begin{equation}
\ubf(\phi,\theta)^{\Her}\Ubf_{\mathrm{n},t}\Ubf_{\mathrm{n},t}^{\Her}\ubf(\phi,\theta)
\label{eq:appB_proj_energy}
\end{equation}
is small when $\ubf(\phi,\theta)$ lies close to the signal subspace and large otherwise. This directly yields the Blind MDL-MUSIC pseudo-spectrum in \eqref{eq:music_detect}, whose dominant local maxima provide the anonymous interference directions $\{(\hat\phi_{k,t},\hat\theta_{k,t})\}_{k=1}^{\hat K_t}$.

After the angular peaks are identified, the remaining task is to estimate the noncoherent powers associated with those directions. Let $\hat{\ubf}_{k,t}=\ubf(\hat\phi_{k,t},\hat\theta_{k,t})$ and define the covariance-domain approximation
\begin{equation}
\widehat{\Rbf}_t-\hat\sigma_t^2\Ibf
\approx
\sum_{k=1}^{\hat K_t} G_{k,t}\hat{\ubf}_{k,t}\hat{\ubf}_{k,t}^{\Her},
\label{eq:appB_alpha_ls}
\end{equation}
where $\hat\sigma_t^2$ is estimated from the smallest eigenvalues of $\widehat{\Rbf}_t$. A nonnegative least-squares fit of \eqref{eq:appB_alpha_ls} then produces the gains $\{\hat G_{k,t}\}$, so the final detector output is the anonymous tuple set $\{(\hat\phi_{k,t},\hat\theta_{k,t},\hat G_{k,t})\}_{k=1}^{\hat K_t}$ used by the noncoherent nulling stage.

\section{Interpretation of the Large-Matrix Nulling Trend}
\label{app:large_matrix_interpretation}

This appendix explains how the large-matrix discussion in Section~\ref{sec:large_matrix} should be interpreted for the practically relevant one-target/one-victim ULA setting. We consider the single-victim specialization of \eqref{eq:sigmod},
\begin{equation}
\Ybf
=
\ubf_1\bbf^{\T}+\Wbf,
\qquad
\bbf=g\sbf,
\label{eq:appC_single_model}
\end{equation}
where $\ubf_1$ is the victim spatial signature, $g$ is the effective uplink sensing coefficient, and $\sbf$ collects unit-power uplink symbols. We define the sensing SNR as
\begin{equation}
\gamma_{\mathrm{S}}
\triangleq
\frac{|g|^2}{\sigma^2}.
\label{eq:appC_gamma_s}
\end{equation}

To connect this model to the rank-one matrix-estimation result in~\cite{esterror}, the effective second-order quantities should be read as
\begin{equation}
\begin{aligned}
\tau_u
&=
\gamma_{\mathrm{S}}\Exp\!\left[|u_{1,m}|^2\right],
&
\tau_v
&=
\Exp\!\left[|s(\ell)|^2\right],\\
\tau_w
&=
\frac{\Exp\!\left[|W_{m,\ell}|^2\right]}{N_{\mathrm{tx}}},
\end{aligned}
\label{eq:appC_tau_map}
\end{equation}
with
\begin{equation}
\beta
=
\frac{T}{N_{\mathrm{tx}}},
\alpha
=
\frac{\tau_w}{\tau_u\tau_v}.
\label{eq:appC_alpha_beta}
\end{equation}
Hence $\alpha=O(\gamma_{\mathrm{S}}^{-1})$, and the linear-estimator overlap from~\cite[Eq.~(5.3)]{esterror} implies, in the informative regime, that
\begin{equation}
\eta
\triangleq
|\widehat{\ubf}_1^{\Her}\ubf_1|^2
\approx
\frac{\left[\beta-\alpha^2\right]_+}{\beta+\alpha}
=
1-\frac{\alpha}{\beta}+O(\alpha^2),
\label{eq:appC_eta_scaling}
\end{equation}
so $1-\eta=O(\gamma_{\mathrm{S}}^{-1})$ as the sensing SNR grows. The first-order approximation $1-\alpha/\beta$ is the same quantity used in the notebook as the ``approx eta'' curve.

The nulling behavior depends on how the victim uplink sensing quality and the victim downlink exposure are coupled. If the same geometry/pathloss trend makes the victim simultaneously easier to sense and more exposed to downlink leakage, it is natural to model the pre-nulling victim INR as
\begin{equation}
\gamma_{\mathrm{I}}
=
c\,\gamma_{\mathrm{S}}
\label{eq:appC_gamma_coupling}
\end{equation}
for some proportionality constant $c>0$. Thus, $\gamma_{\mathrm{S}}$ and $\gamma_{\mathrm{I}}$ are not the same quantity: $\gamma_{\mathrm{S}}$ is the victim uplink sensing SNR, whereas $\gamma_{\mathrm{I}}$ is the pre-nulling downlink victim INR. In the coupled-geometry model above they scale proportionally, but they are not identical in general. The residual leakage after nulling is then
\begin{equation}
\INR
=
\gamma_{\mathrm{I}}
\big|
\widehat{\vbf}^{\Her}\ubf_1
\big|^2.
\label{eq:appC_inr_def}
\end{equation}

To expose the role of the estimation error, we use the results of \cite{esterror} to write 
\begin{equation}
\widehat{\ubf}_1
=
\sqrt{\eta}\,\ubf_1+\ebf,
\qquad
\ubf_1^{\Her}\ebf=0,
\qquad
\|\ebf\|_2^2=1-\eta.
\label{eq:appC_error_decomp}
\end{equation}
Moreover, under the isotropic-error approximation, $\ebf$ is modeled as
zero-mean over the subspace orthogonal to $\ubf_1$, with per-component variance
\begin{equation}
    \Exp |e_i|^2 = \frac{1-\eta}{N_{\subsf rx}}.
\end{equation}
Let $\ubf_0$ denote the desired-user direction, and define
$\rho\triangleq \ubf_1^{\Her}\ubf_0$ and
$\delta\triangleq \ebf^{\Her}\ubf_0$. Then
\begin{equation}
\hat{\rho}
=
\widehat{\ubf}_1^{\Her}\ubf_0
=
\sqrt{\eta}\,\rho+\delta.
\label{eq:appC_rhohat}
\end{equation}
Also, since $\delta\triangleq \ebf^{\Her}\ubf_0$, it will be complex Gaussian with mean zero and variance:
\begin{equation}
    \Exp|\delta|^2 = \Exp |\ebf\herm \ubf_0 |^2 =  \frac{1-\eta}{N_{\subsf rx}} \rightarrow 0.
\end{equation}
The beam $\widehat{\vbf}$ is the unit-norm single-null beamformer obtained by
projecting $\ubf_0$ onto the orthogonal complement of the estimated victim
direction $\widehat{\ubf}_1$:
\begin{equation*}
\widehat{\vbf}
=
\frac{\ubf_0-\hat{\rho}\,\widehat{\ubf}_1}
{\sqrt{1-|\hat{\rho}|^2}},
\qquad
\widehat{\vbf}^{\Her}\widehat{\ubf}_1=0.
\end{equation*}
Therefore,
\begin{align*}
\widehat{\vbf}^{\Her}\ubf_1
&=
\frac{(\ubf_0-\hat{\rho}\,\widehat{\ubf}_1)^{\Her}\ubf_1}
{\sqrt{1-|\hat{\rho}|^2}} \\
&=
\frac{\ubf_0^{\Her}\ubf_1-\hat{\rho}^{*}\widehat{\ubf}_1^{\Her}\ubf_1}
{\sqrt{1-|\hat{\rho}|^2}} \\
&=
\frac{\rho^*-(\sqrt{\eta}\rho^*+\delta^*)\sqrt{\eta}}
{\sqrt{1-|\sqrt{\eta}\rho+\delta|^2}} \\
&=
\frac{(1-\eta)\rho^*-\sqrt{\eta}\,\delta^*}
{\sqrt{1-|\sqrt{\eta}\rho+\delta|^2}}.
\end{align*}
Where $\delta^*$ is the complex conjugate of $\delta$. Taking the squared magnitude gives the residual leakage
\begin{align}
    \Exp\INR &= \gamma_1 \Exp|\widehat{\vbf}^{\Her}\ubf_1|^2 \nonumber \\
    &=\gamma_1 \Exp \frac{|(1-\eta)\rho - \sqrt{\eta}\,\delta^*|^2}{1- |\sqrt{\eta} \rho + \delta|^2}.
    \label{eq:appC_inr_exact}
\end{align}
The passage to the next expression uses a first-order expansion of the
denominator.  Let
\begin{equation*}
x \triangleq |\sqrt{\eta}\rho+\delta|^2 .
\end{equation*}
When $\rho$ and $\delta$ are small, $x\ll 1$, and
\begin{equation*}
\frac{1}{1-x}
=
1+x+O(x^2).
\end{equation*}
Therefore,
\begin{align*}
\frac{|(1-\eta)\rho-\sqrt{\eta}\,\delta^*|^2}
{1-|\sqrt{\eta}\rho+\delta|^2}
&=
|(1-\eta)\rho-\sqrt{\eta}\,\delta^*|^2
\left(1+x+O(x^2)\right) \\
&\approx
|(1-\eta)\rho-\sqrt{\eta}\,\delta^*|^2
\left(1+|\sqrt{\eta}\rho+\delta|^2\right).
\end{align*}
Now we take the approximation where higher-order terms in $\delta$ and
$\rho$ are small, then:
\begin{align}
    \Exp\INR &\approx \gamma_1 \Exp|(1-\eta)\rho - \sqrt{\eta}\,\delta^*|^2(1+ |\sqrt{\eta} \rho + \delta|^2)  \\
    &\approx \gamma_1 \left[ (1-\eta)^2|\rho|^2 + \eta \Exp|\delta|^2 \right] \\
    &= \textcolor{blue}{\gamma_1 \left[ (1-\eta)^2|\rho|^2 + \eta \frac{1-\eta}{N_{\subsf rx}} \right]}.
\end{align}

Consider two special cases.  First, if there is perfect estimation, $\eta = 1$ and $\INR = 0$
which makes sense we perfectly null the victim.  On the other hand, if there is worst-case error, $\eta = 0$ and 
\[
    \INR = \gamma_1 |\rho|^2.
\]
This value is the same as if the transmitter performed no nulling.

Equation \eqref{eq:appC_inr_exact} shows that there are two distinct regimes.
If the error component along the desired-user direction is asymptotically negligible, i.e., $\delta=o(\sqrt{1-\eta})$, then $|\widehat{\vbf}^{\Her}\ubf_1|^2=O((1-\eta)^2)$, and with \eqref{eq:appC_eta_scaling} and \eqref{eq:appC_gamma_coupling} the residual INR eventually decreases like $O(\gamma_{\mathrm{S}}^{-1})$. This is the optimistic regime behind the simplified isotropic-error approximation.

In contrast, in a finite-dimensional ULA system it is common that the estimation error still has a non-negligible projection on $\ubf_0$, so $\delta=O(\sqrt{1-\eta})$. Then \eqref{eq:appC_inr_exact} yields $|\widehat{\vbf}^{\Her}\ubf_1|^2=O(1-\eta)$, and therefore
\begin{equation}
\INR
=
\gamma_{\mathrm{I}}\,O(1-\eta)
=
c\,\gamma_{\mathrm{S}}\,O(\gamma_{\mathrm{S}}^{-1})
=
O(1).
\label{eq:appC_inr_plateau}
\end{equation}
That is, the residual INR no longer grows without bound; instead it approaches a finite ceiling.

When the victim uplink is weak, the GLRT estimate $\widehat{\ubf}_1$ is inaccurate and the estimated null is poor; as the victim becomes stronger, the raw interference level also increases, so the post-nulling INR initially rises; once the sensing SNR is high enough that $\eta$ is close to one, the improvement in $\widehat{\ubf}_1$ offsets the additional raw interference, and the residual INR becomes approximately flat. In other words, the qualitative trend is ``increase first, then saturate,'' not indefinite growth. The notebook therefore compares the Monte Carlo nulling curve against \eqref{eq:appC_inr_exact} using both the exact overlap model in \eqref{eq:appC_eta_scaling} and its first-order approximation $1-\alpha/\beta$.

Therefore, the main message of Section~\ref{sec:large_matrix} is qualitatively correct for the coupled uplink/downlink setting that motivated the notebook validation: stronger victim uplink improves GLRT estimation, and this eventually limits the residual nulling leakage. What is directly inherited from~\cite{esterror} is the overlap scaling in \eqref{eq:appC_eta_scaling}; the finite nonzero INR ceiling in \eqref{eq:appC_inr_plateau} is a practical finite-dimensional consequence of coupling $\gamma_{\mathrm{I}}$ with $\gamma_{\mathrm{S}}$ and of retaining the residual error component $\delta$ along the desired-user direction.

\end{document}